\documentclass[]{IEEEphot}

\jvol{xx}
\jnum{xx}
\jmonth{Mar}
\pubyear{2020}
\usepackage[dvipsnames]{xcolor}
\usepackage{amsmath}
\usepackage{comment}
\usepackage{subcaption}
\usepackage{soul}
\usepackage{caption,subcaption}
\usepackage[labelformat=parens,labelsep=quad,skip=3pt]{caption}
\usepackage{graphicx}
\usepackage{caption,subcaption}

\usepackage{cite}
\usepackage{gensymb}
\usepackage{ulem}
\usepackage{textcomp}
\begin{document}

\title{Optical Characterization of Ultra-Low latency Visible Light Communication System for Intelligent Transportation Systems}

\author{M. Seminara\IEEEauthorrefmark{1}, T. Nawaz\IEEEauthorrefmark{2}\IEEEauthorrefmark{4}, S. Caputo\IEEEauthorrefmark{3}, L. Mucchi\IEEEauthorrefmark{3}~\IEEEmembership{Senior~Member,~IEEE}, \\ and J. Catani\IEEEauthorrefmark{4},\IEEEauthorrefmark{1}}

\affil{\IEEEauthorrefmark{1} European Laboratory for NonLinear Spectroscopy (LENS), University of Florence, Sesto Fiorentino, Italy. \\ \IEEEauthorrefmark{2} Dept. of Physics and Astronomy, University of Florence, Italy. \\
\IEEEauthorrefmark{3} Dept. of Information Engineering, University of Florence, Italy. \\  
\IEEEauthorrefmark{4} Istituto Nazionale di Ottica del CNR (CNR-INO), Sesto Fiorentino, Italy.}

\doiinfo{DOI: ZZZZZZZ/JPHOT.YYYY.XXXXXXX\\
XXX-XXXX/\$25.00 \copyright 2020 IEEE}%

\maketitle

\markboth{IEEE Photonics Journal}{YYY}
\thanks{\textcopyright 2019 IEEE. Personal use of this material is permitted.  Permission from IEEE must be obtained for all other uses, in any current or future media, including reprinting/republishing this material for advertising or promotional purposes, creating new collective works, for resale or redistribution to servers or lists, or reuse of any copyrighted component of this work in other works.}
\vspace{10mm}

\begin{receivedinfo}%
Manuscript received XXXX. Corresponding author: jacopo.catani@ino.cnr.it
\end{receivedinfo}

\begin{abstract}

This paper reports a detailed experimental characterization of optical performances of Visible Light Communication (VLC) system using a real traffic light for ultra-low latency, infrastructure-to-vehicle (I2V) communications for intelligent transportation systems (ITS) protocols. Despite the implementation of long sought ITS protocols poses the crucial need to detail how the features of optical stages influence the overall performances of a VLC system in realistic configurations, such characterization has rarely been addressed at present. 
We carried out an experimental investigation in a realistic configuration where a regular traffic light (TX), enabled for VLC transmission, sends digital information towards a receiving stage (RX), composed by an optical condenser and a dedicated amplified photodiode stage. We performed a detailed measurements campaign of VLC performances encompassing a broad set of optical condensers, and for TX-RX distances in the range 3--50\,m, in terms of both effective field of view (EFOV) and packet error rate (PER). The results show several nontrivial behaviors for different lens sets as a function of position on the measurement grid, highlighting critical aspects as well as identifying most suitable optical configurations depending on the specific application and on the required EFOV. In this paper we also provide a theoretical model for both the signal intensity and the EFOV as a function of several parameters, such as distance, RX orientation and focal length of the specific condenser. To our best knowledge, there are no optical and EFOV experimental analyses for VLC systems in ITS applications in literature. Our results could be very relevant in the near future to assess a most suited solution in terms of acceptance angle when designing a VLC system for real applications, where angle-dependent misalignment effects play a non-negligible role, and we argue that it could have more general implications with respect to the pristine I2V case mentioned here. 
\end{abstract}

\begin{IEEEkeywords}
{Visible light communication, intelligent transportation systems, optical systems, infrastructure to vehicle, vehicle to vehicle, field of view,  channel model.} 
\end{IEEEkeywords}

\section{Introduction}


Visible Light Communication (VLC) technology is envisioned as one of the most favourable candidates for efficient implementation of complex Intelligent Transportation Systems (ITS) protocols, aimed at improving safety an efficiency of the urban mobility scenario \cite{VLCINITS7} by introducing a pervasive wireless data inter-exchange between infrastructures and vehicular units (I2V) and between vehicles (V2V). VLC technology exploits light in the visible spectrum [400-750\,nm] as optical carrier to transmit digital data wirelessly \cite{PHY2} \cite{IEEE802.15.7}. After the commercial spread of high-power LED sources, VLC has proven to grant for either high data rates for indoor wireless applications (Li-Fi) \cite{PHY1}, or pervasive broadcast of short information packets with very low latency, which is especially important in outdoor ITS safety-critical applications \cite{our_paper2} where smart vehicles must be equipped with ultra-reliable and low-latency communication systems to share information with infrastructures and nearby vehicles for triggering autonomous or assisted actions aimed at avoiding critical events.

One of the key features of the VLC technology in ITS applications relies on the intrinsic directionality of the optical channel. In a VLC system, optical elements such as lenses or compound parabolic concentrators \cite{CPC, Ning:87} can be used both at transmitter side to provide for light-beam shaping capabilities, and at receiver side, as image-forming stages in camera-based implementations \cite{VLCSENSOR12,VLCSENSOR13,SURVEY85l,SURVEY86,SURVEY87}, or as optical antennas in photodiode-based schemes \cite{SURVEY60,SURVEY64,SURVEY69,SURVEY70,SURVEY71,SURVEY72,SURVEY89,our_paper2}, to increase the optical gain of the optical detector, hence improving communication distances and link quality at the expense of reducing the angle of field of view (AFOV) of receiver.

In order to enable fast and robust vehicular communications several technologies and techniques \cite{Vehicularcomm} have been proposed and tested, but most efforts focused on dedicated on RF-based short range communications (DSRC) and IEEE standard 802.11p, which forms the regulations for wireless access in vehicular environments (WAVE) \cite{DSRc,Hrdprototype5}. These standards use dedicated  frequency  bands  for  ITS in Europe and United States to provide the potential solutions for future implementations of communication-based ITS safety applications \cite{VLCINITS3,VLCINITS4,Hrdprototype}.  
\color{black}

In this scenario, the directionality of optical channel can offer several advantages with respect to more established radio-frequency (RF)-based wireless communication technologies, such as the intrinsic degree of security in indoor applications (where light can be more easily confined by walls or doors with respect to RF fields, typical of e.g. Wi-Fi or mobile technologies) or the possibility to avoid complex network architectures and packet structures thanks to the strict  line-of-sight (LoS) nature of the optical links \cite{PHY1}. The latter feature, in association to a very high degree of integrability of VLC in forecoming mobile technologies such as 5G \cite{5GV2X}, offers a very important opportunity especially in outdoor ITS I2V and V2V applications, where the implementation of reliable and effective safety and autonomous driving  protocols raises the need for ultra-low latency wireless data delivery \cite{5Glatency}. In view of real ITS implementations of VLC networks, one has anyhow to consider that the relative positions of vehicles and infrastructures is typically bound to road geometries and signaling infrastructures regulations (such as height and position on the road), and the ideal LoS condition where optical axes of TX and RX optical equipments coincide is far from being a valid approximation. In such realistic conditions the angular misalignment between TX and RX units can achieve large values, with strong implications on the quality of the VLC communication channel \cite{7980780}, as well as the achievable haul of the VLC link. In this context, whilst a wealth of theoretical and experimental efforts have been made in recent years on development of VLC systems and protocols for ITS I2V and V2V scenarios \cite{Survey_VLC_ITS}, a detailed study of VLC setups in terms of optical performances, and in particular, the characterization of performances of a VLC link for different optics sets in terms of communication cast and effective Field of View of the system, would be of primary importance to assess the range of applicability of ITS protocols in realistic vehicular implementations. 
\color{black}

In this paper we report a thorough optical characterization of a newly-designed VLC system for I2V ITS applications, based on low-cost, open-source digital platforms and exploiting a real traffic light \cite{our_paper2}. We perform an exhaustive experimental campaign to determine the optical and transmission properties of our system for a wide set of commercial aspherical and Fresnel optical condensers at RX stage with different diameters and focal lengths. We characterize the performances of the VLC link in realistic configurations, for several positions on the measurement grids and for distances up to 50\,m by varying the RX stage orientation relative to the traffic light lamp and measuring an effective Field of View (EFOV) in terms of Packet Error Rate (PER) and received amplitude. We characterize limits and advantages of each lens, quantifying the dramatic effects of relative angular misalignment between TX and RX stages at short distances. We propose the most suitable optical configurations depending on the specific application, finding that our recent VLC system \cite{our_paper2} allows for error-free communications at 115\,kbaud for distances up to 50\,m, with measured EFOVs higher than 10\degree\, when equipped with aspheric lenses. We also present an optical model for received intensity and EFOV, which is an angle-dependent extension of the intensity model given in \cite{2019arXiv190505019C}, featuring an excellent predictive capacity when compared against data. This model has actually more general outcomes with respect to the specific case of I2V reported in this work, as it quantifies an intrinsic performances of a VLC RX stage, composed by an optical condenser of a given focal length and a photodiode of a give size and can be of help predicting the angle-dependent performances of virtually any photodiode-based VLC system once the intensity map emitted by the TX source is know.
The optical characterization reported in our work represents a strong advance towards the integration of VLC technology in real ITS applications, where the characterization of performances of the optical link in presence of non-ideal and finite-size effects is essential.

The rest of the paper is organised as follows:
Section \ref{sec:setup} describes the VLC system and the  experimental setup for its optical characterization, as long as the optical elements employed. Section \ref{sec:model} describes an optical model, aimed at predicting the expected FOV of our VLC system given the intensity map of the real source. The experimental results on characterization of optical and transmission performances in terms of field of view are then presented and discussed in Section \ref{sec:Exp_results}, before we conclude this paper in Section \ref{sec:conclusions}.

\section{Experimental Setup} \label{sec:setup}

\begin{figure}[]
\centering
\includegraphics[width=0.65\columnwidth]{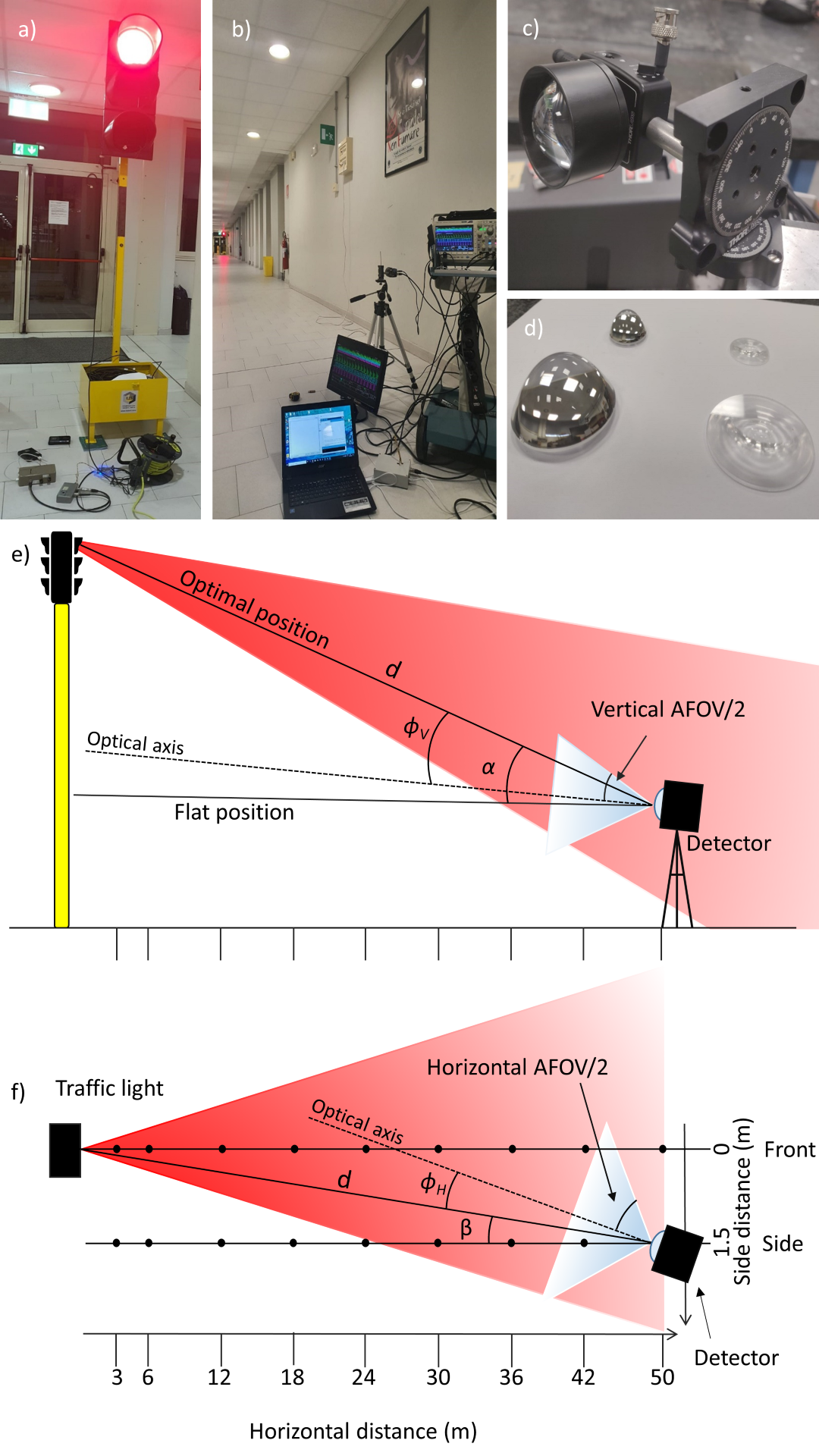}
\caption{Experimental setup: a) TX unit: a standard traffic light, digital transmitter board and current modulator. b) Receiver unit: photodetector, PC, oscilloscope and digital RX board. c) Photodetector with variable gain and rotational platforms. d) Lenses, namely 1" and 2" Fresnel and aspherical. e)-f) Schematic side and top views of the measurement grid. The solid lines in e) highlight the two optimal and flat configurations for RX optical axis orientation, whilst solid-dotted lines f) highlight the two horizontal configurations used (front, 1.5\,m side, respectively) across the measurement grid. The TX-RX distance  is  up  to  50  m  limited  by  available LoS length in the building.}
\label{fig:Exp_setup}
\end{figure}

The experimental campaign is aimed at optical characterization of performances of a recent low-cost, open-source, ultra-low latency VLC link for ITS communications \cite{our_paper2}, and in particular of its I2V branch. We perform a thorough analysis of received signal amplitude and corresponding PER as a function of distance and receiver-traffic light relative orientation and position in order to extract relevant information about the Angle of Field of View (AFOV) of the receiver. We perform the campaign in a realistic scenario and for several condensing lenses, to experimentally determine the effect of lens typologies (aspherical/Fresnel), diameters, and focal lengths on the communication performances of the VLC system.
Experiments are performed in a 60 m-long corridor in the Department of Physics and Astronomy at University of Florence (see Fig.\ref{fig:Exp_setup}), where both sunlight and artificial lights were present. As already proven previously  \cite{2019arXiv190505019C} the DC sunlight component is efficiently rejected by our RX system even in outoor scenarios. On the other side, as we will see later on, residual low-frequency effects on optical link due to artificial illumination 
leading to possible false-trigger events in comparator induced by residual 100 Hz background noise, and to secondary ceiling and floor reflections of the traffic light signal, are promptly isolated by our measurement procedure. Despite measurements are taken in an indoor environment, hence, our results provide for valuable information on performances and FOV of a VLC link in outdoor real scenarios which, in case of absence of heavy rain or fog (to be analyzed in future works), are rather representing an "easier" scenario in terms of communication performances.

\subsection{Transmitter and receiver stages}\label{sub:TXRX}
Our experimental setup, along with side and top views of the measurements grid, is shown in Fig.~\ref{fig:Exp_setup}. 
As details on our VLC prototype have already been given elsewhere \cite{our_paper2}, and being them unnecessary to the scope of the present work, we only give a brief resume of the VLC system architecture. The transmitter unit (TX, Fig.\ref{fig:Exp_setup}a) is composed by a digital encoder/modulator which is realised through a microcontroller-based digital board (Arduino DUE), which provides a digital modulation signal, fed into a regular traffic light LED lamp (red) by means of a proprietary-design analog high-current modulation stage. The traffic light has been provided by ILES srl, a company for smart signaling solutions based in Prato (Italy). This module is able to transmit a continuos stream of data packets (broadcast) with a maximum data rate of 230 Kbps. The TX uses On-Off Keying (OOK) modulation with Non-Return-to-Zero (NRZ) data coding \cite{OOK}. For the present PER analysis, we typically transmit 100k, 6 byte-long packets, through direct modulation of the LED current using the UART protocol provided by Arduino DUE serial ports.
The Receiver unit (RX, Fig.\,\ref{fig:Exp_setup}b-c) is placed at a height corresponding to the car dashboard height of 105 cm. It includes an optical lens providing optical gain by focusing the incoming light on a commercial large-area photodiode with variable gain (Thorlabs PDA36A2), which has been decoupled from DC component through a hardware modification in order to provide for total rejection of the effect of direct sunlight exposure, and more than 30 dB attenuation on the 100-Hz artificial illumination components \cite{our_paper2} (see also Sec.\,\ref{sec:Exp_results}). After the reception and amplification stage, the analog signal is passed through a variable-threshold comparator for re-digitisation. This digital signal is then analysed and decoded by a digital receiver board (based on Arduino DUE). The decoded message is compared against a stored reference message, so that the PER can be measured, along with the amplitude of the received signal, which is recorded by means of a 1 Gs/s 4-channel digital oscilloscope, allowing us to measure the amplitude of the received signal after complete subtraction of the low-frequency 100 Hz-component in post-analysis. This fundamental capability allows us to accurately determine the performances of our system in a configuration where 100-Hz components are absent, like in outdoor applications (see Sec.\,\ref{sec:Exp_results}). The RX stage is mounted on a tip/tilt platform stage granting 0.5\textdegree of resolution in the horizontal and vertical orientation angles of detector ($\beta$ and $\alpha$, respectively, see Fig. \ref{fig:Exp_setup} e-f).

\subsection{Condenser lenses}\label{subsec:lenses}
 Achieving at the same time large AFOV values and long hauls would represent the main target of any VLC system in ITS applications. However, this task would require lenses with very large diameters and very short focal lengths, which can be problematic both in terms of feasibility, aberrations, costs, and size of RX unit in vehicular applications. Therefore, we restrict our analysis to off-the-shelf, low-cost (5 to 40 Euros), 1" and 2" diameter lenses with shortest available focal lengths (see Tab. 1), as they are among the most suitable non-custom available candidates as condensing optical elements in ITS applications. We analyzed both molded glass aspheric lenses (more expensive, better optical performances) and plastic Fresnel ones (less expensive, lower thickness and weight, but typically larger optical aberrations due to their finite-element surface structure). We argue that in our RX system, when large angles between TX and RX stages are involved (see Fig.\,\ref{fig:Exp_setup}e-f) and Fig.\,\ref{fig:Model_geometry}) the most relevant aberration affecting the effective capability to collect a signal coming from off-axis sources is coma, whilst spherical aberration, chromatism, field curvature and image distortion will have a marginal impact on performances as a non-imaging light detection is only required our photodiode-based VLC setup.
 
\begin{table}
\begin{center}
\begin{tabular}{|c|c|c|c|c|} \hline 
 Diameter & Focal length & Type & Vendor/code & Acronym  \\ \hline
 1" & 16 mm & Aspheric Molded Glass & Thorlabs ACL25416U & AS1\\
 2" & 32 mm & Aspheric Molded Glass & Thorlabs ACL50832U & AS2\\
 1" & 25 mm & Fresnel Plastic  & Thorlabs FRP125 & FR1\\
 2" & 32 mm & Fresnel Plastic  & Thorlabs FRP232 & FR2\\\hline 
\end{tabular}\end{center}\label{table:lenses}
\caption{TABLE 1: Lens set used in the experiments.}
\end{table}


\section{Optical and Channel Model} \label{sec:model}
\begin{figure}[t]
 \centering\includegraphics[width=\columnwidth]{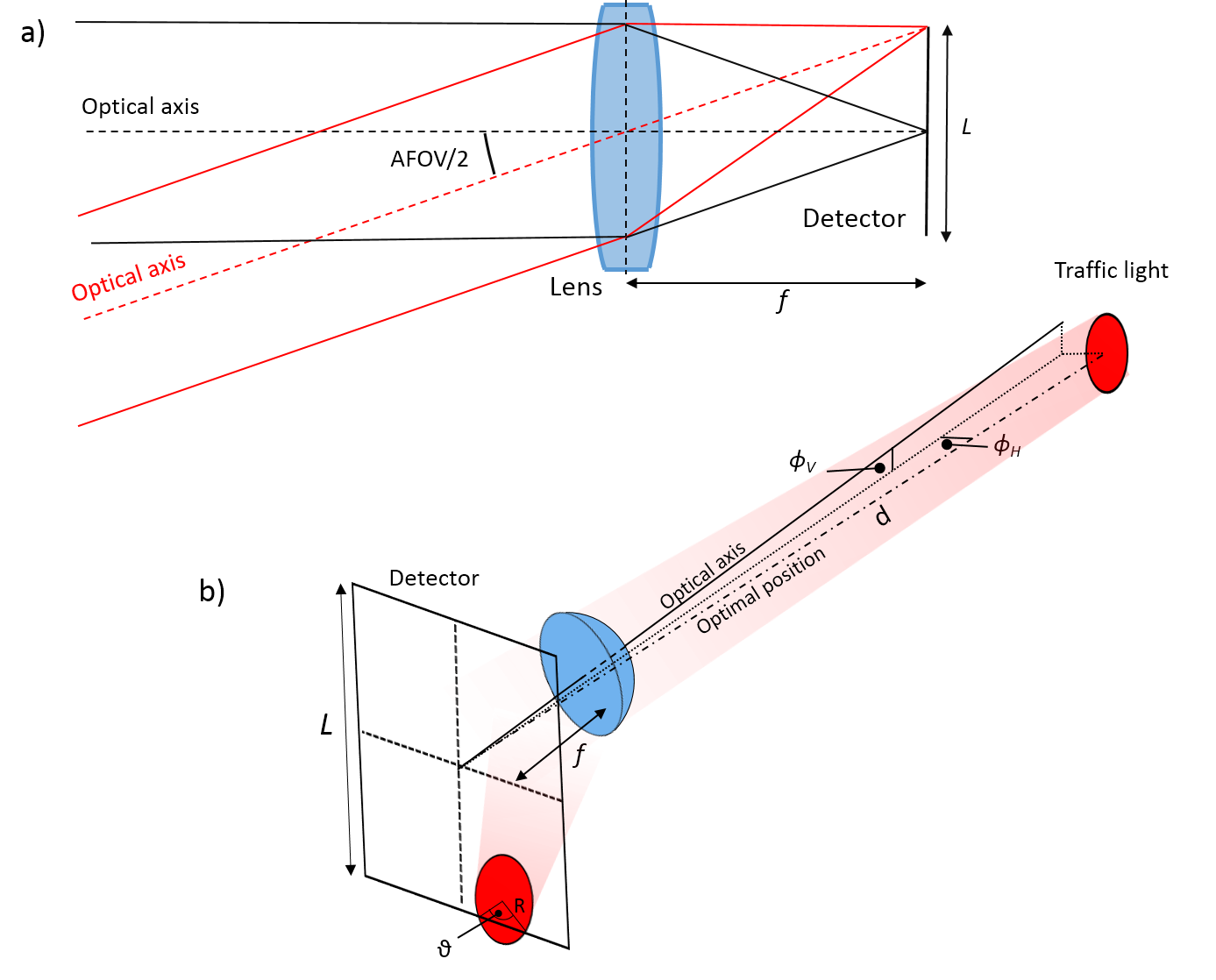}
\caption{a) Off-axis ideal source at infinity: the AFOV depends on the condenser focal length $f$ and on dimension of the photodetector $L$. b) Extended source (homogeneous disk): the AFOV depends also on the dimension of the image $R$. The portion of image falling inside the photodiode area is parametrized in terms of angle $\theta$, which is correlated with the angles $\phi_{H,V}$, see Fig\ref{fig:Exp_setup} e-f.}
\label{fig:Model_geometry}
\end{figure}

The model presented in this section aims at the description of the optical power, i.e. the signal amplitude, measured at RX in our VLC system, as a function of relative TX-RX orientation and position on the grid. Indeed, this is the most important parameter determining the performances of a VLC system in realistic applications and allows us to perform a characterization of the AFOV in terms of quality of our VLC link. Starting from results of a previous work \cite{2019arXiv190505019C}, where a model for intrinsic intensity map $I(\alpha,\beta, d)$ for the same traffic light as a function of elevation angle $\alpha$, azimuth $\beta$ and TX-RX distance $d$ (see Fig.\,\ref{fig:Exp_setup}e-f) is given, we construct a more exhaustive model where the explicit dependence of received signal amplitude (hence not only intensity) as a function of relative angular orientation of TX (traffic light) and RX is obtained. We use model $I_1$ proposed in \cite{2019arXiv190505019C} since it was the one providing for the best trade-off between accuracy and complexity. \color{black}
 
For sources at infinite distance (see Fig.\,\ref{fig:Model_geometry}) AFOV is defined as the angular displacement of source object relative to the optical axis for which the image is still entirely formed into the detector active area, and depends on lens focal length  $f$ and on half-size $L/2$ of detector active area \cite{edmundwebsite}:
$
\mathrm{AFOV}= 2\,\mathrm{tan}^{-1}(L/2f).
$

\begin{figure}[!t]
\begin{center}

\includegraphics[width=\columnwidth]{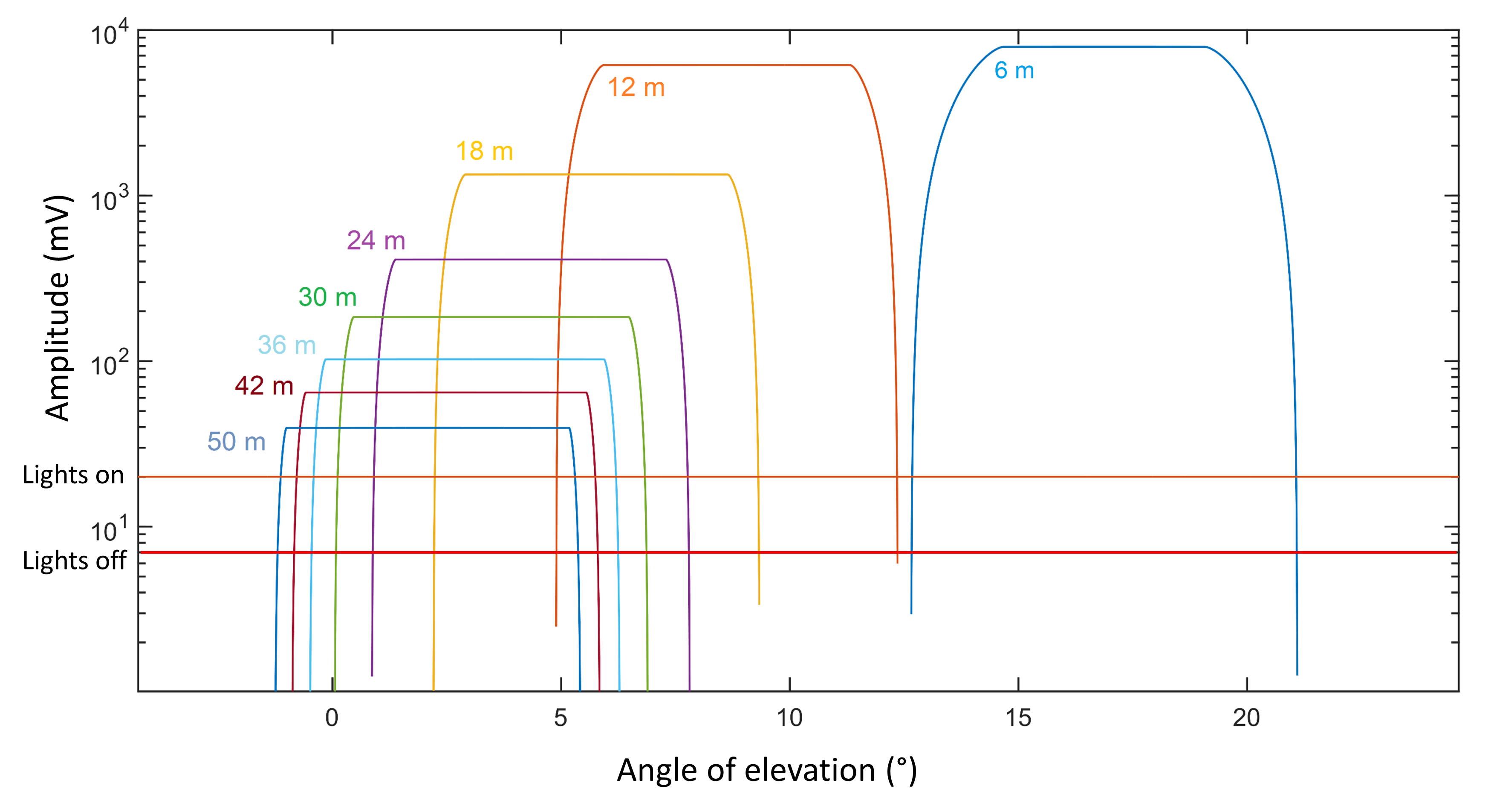}
\caption{Amplitude of received signal, calculated by our model as a function of angle of elevation $\alpha$ for AS2 lens for various distances as measured in the optimal configuration (see text).The horizontal lines show the EFOV (PER=$10^{-3}$ when artificial illumination is on (brown) and and off (red) (Sec. \ref{sub:trans_performances})}
\label{fig:Mode_asp}
\end{center}
\end{figure}

However, for extended sources, and in presence of non idealities of the optical system (e.g. aberrations), the effective amount of light collected on photodiode, as the RX is tilted, is not strictly given by AFOV as it depends on the shape and size of the image of the source.  By assuming our source as a circular disk, and neglecting aberrations for now, we can estimate the radius of the image $R$ by using standard thin lens equation for all employed lenses and distances of interest. Assuming the source as a uniform 30 cm-diameter disk, we compute $R$ to be always smaller than $L$, for all distances of interest (6--50\,m). Three different cases can hence arise when measuring the optical power detected by the receiver as its orientation is varied: 1) image is entirely formed  inside the photodiode; 2) image is formed completely outside the photodetector and 3) only part of image is formed inside the photodiode. In the third case an angle $\theta$ is used to parametrize both inner and outer areas of image (See Fig.\,\ref{fig:Model_geometry}-b). The outside region is a circular segment of area  $A=\frac{R^2}{2}(\theta-\mathrm{sin}(\theta))$. For sake of simplicity, and with no loss of generality, we assume now to aim the optical axis of receiver at the traffic light lamp ("optimal" position of Fig.\,\ref{fig:Exp_setup}e-f), and to rotate the RX only along one its vertical axis (the calculation is equivalent for rotation around the horizontal axis). The angle $\theta$ is correlated to the angle $\phi_V$ between photodiode and traffic light:
\begin{equation}
\theta=2\,\mathrm{cos}^{-1}\left(\frac{L/2-f\,\mathrm{tan}(|\phi_V|)}{R}\right).
    \label{thetaeq}
\end{equation} 
 Then, the amplitude $S(\alpha, \beta, d)$ of VLC signal provided by photodiode to the digital RX board, which is proportional to the intensity of light collected by photodiode through a calibration factor $c$, depends on the area of the image formed inside the photodiode and it is given by:
\begin{equation}
  S(\alpha, \beta, d)=c\cdot\begin{cases}
    0, & \text{$-\frac{\pi}{2}<\phi_V\leq -\phi_2$ $\vee$ $\phi_2\leq\phi_V<\frac{\pi}{2}$},\\
    I(\alpha,\beta,d)cos(\phi_V) & \text{$-\phi_1\leq\phi_V\leq\phi_1$},\\
   I(\alpha,\beta,d)cos(\phi_V)(1-\frac{\theta-\mathrm{sin}(\theta)}{2\pi}) & \text{$-\phi_2<\phi_V< -\phi_1$ $\vee$ $\phi_1<\phi_V\leq\phi_2$},
  \end{cases}
  \label{eq:model_eq}
\end{equation}

The angles $\phi_1=\tan^{-1}\left( \frac{L/2-R}{f} \right)$ and $\phi_2=\tan^{-1}\left( \frac{L/2+R}{f} \right)$ define a region which, in contrast to the ideal case of point-like image mentioned above, correspond to a non-sharp transition between maximum and zero collected power, which measure the maximum effective angle by which the Rx can be misaligned before the signal is lost. A $\phi$-dependent amplitude map can then be retrieved from intensity map $I(\alpha,\beta,d)$ predicted by model once the coefficient $c$ is measured on a grid's point. Fig. \ref{fig:Mode_asp} shows the simulated amplitude map, as a function of elevation angle $\alpha$, for AS2 lens, when the receiver is rotated in vertical direction. Due to symmetry considerations on RX stage, we expect the same behaviour also for the horizontal case. The calibration factor $c$ appearing in model given above is obtained by normalizing the maximum amplitude $I(\alpha, \beta, d)$ at a distance of 6 m to the absolute measured value in this position. In order to retrieve information on telecom performances of the VLC setup, we define an effective field of view (EFOV) as the angle $\phi_{H,V}$ below which the observed PER is better than $0.1\%$  which is the most common recommended threshold for high reliable communication applications \color{black} (represented by horizontal lines of Fig.\,\ref{fig:Mode_asp} with and without artificial light interference; see Sec.\, \ref{sub:trans_performances} for details on experimental calibration).
The observed EFOV spans the range $\sim$ 5--10 degrees and reduces as RX stage is moved away from traffic light. As curves show, the transition region $[\phi_1, \phi_2]$ gets narrower as the distance is increased. This is mainly due to a larger demagnification of optical system at larger distances, where the image size is smaller. For the same reason, the first case of Eq.\,\ref{eq:model_eq} is favored at large distances, as the larger plateaus show, making it easier to form an image inside the active area of RX. The reduced intensity at larger distances, however, is the dominant effect, so that the EFOV globally shrinks as the distance increases, making it harder to achieve good connections as RX is misaligned.

\section{Experimental results and discussions}  \label{sec:Exp_results}

\begin{figure}[!t]
\centering\includegraphics[width=\columnwidth]{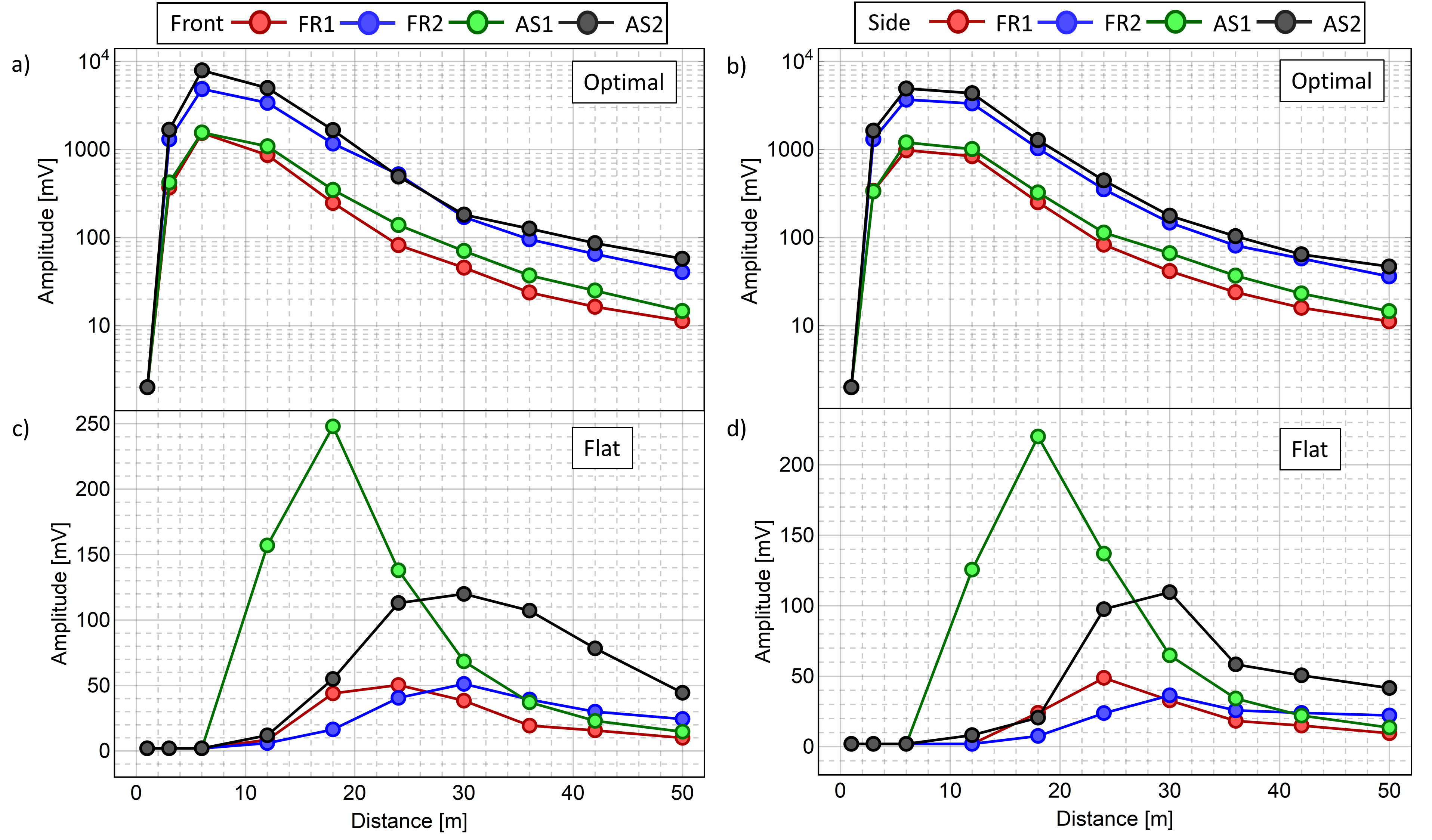}

  \caption{ Amplitude of received signal as a function of distance for four optical lenses ( 1" AS, 1" FR, 2" AS, and 2" FR ) and two receiver positions front (aligned with transmitter) and side (1.5 m from alignment position). The upper panel shows the optimal configuration in which receiver is always point towards the transmitter and lower panel shows the flat case where receiver is parallel to floor.}
\label{fig:Amp_vs_distance}
\end{figure}

For optical characterization of our I2V VLC system, we evaluate performances of the overall system, for all of the 4 lenses described in Sec. \ref{subsec:lenses}, in terms of PER and AFOV for several different positions on the measurement grid (see points of Fig.\,\ref{fig:Exp_setup}-f), reaching distances up to 50\,m. The detector is either placed in line to the traffic light or on a lateral line, displaced by 1.5\,m, in order to analyze a realistic configuration where a car is approaching a crossing along a road following the rightmost lane. The FOV analysis is performed by recording the signal amplitude during a scan of angles $\phi_H$ and $\phi_V$. The angular configurations chosen for both the "front" and "side" settings are two. In a first configuration ("optimal" condition, see Fig. \ref{fig:Exp_setup}) we align the RX axis with the center of the LED lamp, which corresponds to a situation where the car's RX system is actively tracking the lamp position. In the second ("flat") condition the RX optical axis is horizontal, and parallel to the longitudinal axis of corridor, in order to simulate a realistic situation where the car's RX system is oriented towards the direction of motion with no angular adjustment.

\subsection{Optical performances}\label{sub:optical_performances}
Fig.\,\ref{fig:Amp_vs_distance} a-b shows amplitude recorded in the optimal configuration for front and side cases. In this configuration, the recorded amplitude directly reflects the intensity pattern projected by the traffic light as $\phi_V$ (and $\phi_H$) are 0\textdegree, in Eq. \ref{eq:model_eq}. Two main trends are evident. The first feature is that the signal globally follows the same trend, for all lenses, with an initial fast rise at short distances $<$ 10 m and a following slower decay for larger distances. This is inherently tied to the orientation of intensity pattern emitted by the traffic light lamp \cite{2019arXiv190505019C}, which in combination to the chosen height of RX (105 cm) delivers a very small intensity for short distances whilst showing a maximum around 5--10 m. Second, whilst the average level is depending on the lens used, the global trend is the same for all lenses. In the optimal configuration, indeed, the received power at a given distance only depends on the lens diameter $D$ with a $D^2$ dependence given by the energy flux entering the system's pupil (lens). Interestingly enough, hence, the observed amplitude ratio between 2" and 1" lens sets is practically constant and equal to $\sim 4$ in the whole grid, with no dependence on focal length or kind of lens used. For practical applications, hence, this first analysis suggests a 2" Fresnel lens as the most (cost- and weight-) effective optical concentrator for I2V VLC applications.
The situation is anyhow more complex if the side configuration is considered (Fig.\,\ref{fig:Amp_vs_distance} c-d), where large relative angles between TX and RX are involved especially at short distances. As a consequence, the amplitude average values are globally lower than in the front case, and the trends are strongly dependent on the lens used. At short distances, the large values of $\phi_{H,V}$, in association to the small levels of intensity map for large $\alpha$ and $\beta$, drastically reduce the detected optical power to critical levels. In this situation, AS1 lens, which is the shortest focal length element in the set, outperforms all other leneses for both front and side configurations for short to medium distances up to 25 m, despite the 1" diameter. This behavior is due to the fact that larger AFOV values require smaller focal lengths. However, for longer distances, where the FOV is less critical, the larger optical gain of 2" lenses recovers the behavior of the front case, allowing for a better collection of optical power with respect to 1" lenses. Therefore, in this case it is hard to consider a single lens as most suited for whole transmission range. We will detail in next sections the effects of such behaviour in the transmission performances.

\subsection{Transmission performances}\label{sub:trans_performances}
\subsubsection{PER measurement and analysis}\label{subsub:trans_perf_PER}

PER is an important metric, extensively used to assess performances of telecom links.
Assuming an uniform error distribution, PER can directly be related to bit error rate (BER) $PER= 1-(1-BER)^N$, where N is the length of packet \cite{1429989}. Since as PER detection is typically less demanding than BER in terms of real-time signal processing power, we considered PER as a performance metric for our system. Notwithstanding such assumption, a statistically-relevant PER/BER measurement involves long acquisition times, undermining the practical possibility to perform a full characterization of PER for various lenses and as a function of the RX orientation angles throughout a sizeable measurement grid. In order to perform such characterization, we notice that in our Amplitude Shift Keying (ASK) OOK protocol, in turn, BER (hence PER under above assumptions) has a direct relation to the amplitude of the received signal through the Q-function, $BER= Q(\sqrt{\mathrm{SNR}})$ \cite{digitalcomm}, where SNR is the signal-to-noise ratio.  

\begin{figure}[]
\centering\includegraphics[width=\columnwidth]{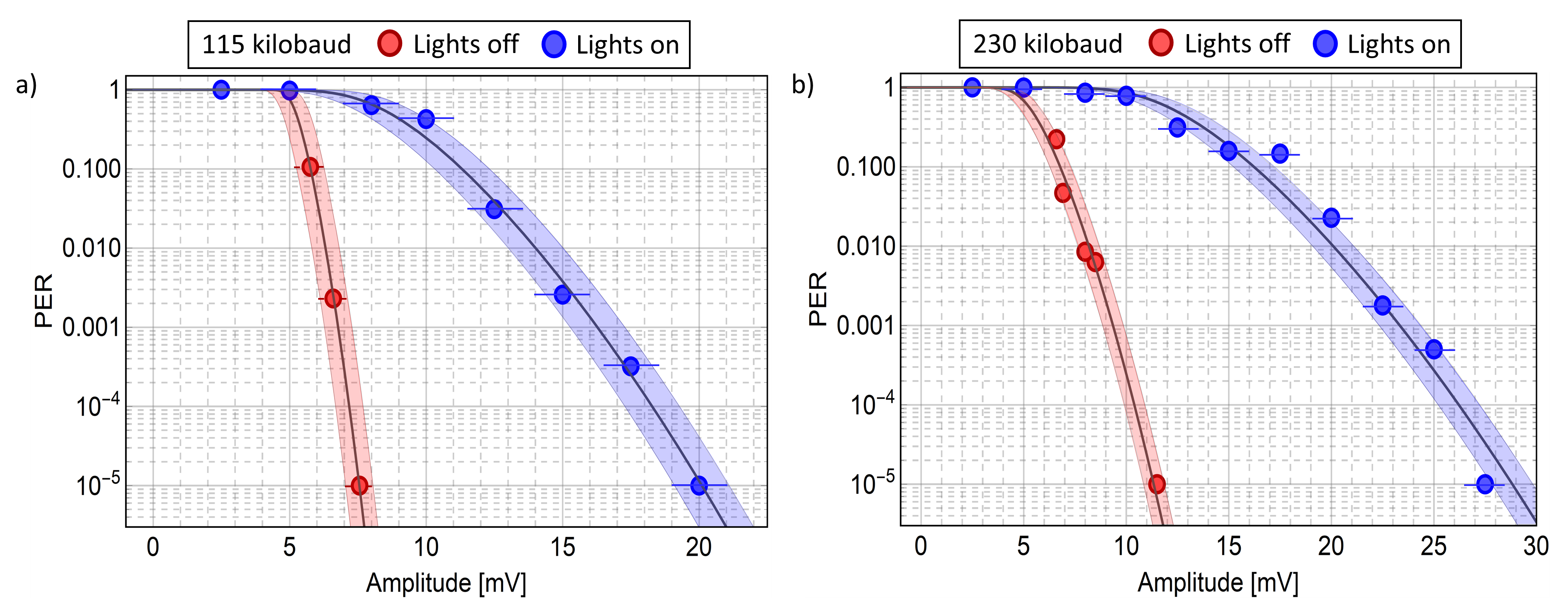}
\caption{PER measurements for various received signal amplitudes. Two baud rates of 115 kbaud and 230 kbaud are used for data transmission. The red dots are the measured values with no artificial lights, while the blue dots are the values when lights are on. Symbols are the average of three consecutive measurements (error bars are masked by symbols). The black curves are fits to data. A statistical error on amplitude values of $\pm$0.5 mV ($\pm$ 1 mV) is considered for the lights-off (lights-on) case. The shaded areas show the error associated to the weighted fit procedure.}
\label{fig:PER_vs_amp}
\end{figure}

We can hence reasonably assume that, given a certain baudrate, the measured PER only depends on detected signal amplitude through the relation
\begin{equation}\label{eq.Qfunc}
    PER= 1-(1-Q(\sqrt{SNR}))^N= 1-\left(1-\frac{1}{2} \text{erfc}\left(\frac{S(\alpha,\beta,d)-T}{\sqrt{2}\sigma}\right)\right)
\end{equation}
where $S(\alpha,\beta,d)$ is the amplitude of the VLC signal and $\sigma$ quantifies the standard deviation of the background noise of the system. The parameter $T$ takes to account the presence of an eventual threshold in the RX due to the comparator stage, designed to avoid false-triggering due to background noise at the expense of introducing a non-zero threshold in the minimum detectable signal level.
Such relation allow us to characterize the PER as a function of amplitude in a single, independent measurements run first, and then use such curve as a calibration for the whole measurements grid in order to retrieve a complete characterization of PER from signal amplitude measurements, by using Eq.\,\eqref{eq.Qfunc} as interpolation function of the experimental data set.  Fig.\,\ref{fig:PER_vs_amp} reports such calibration, for both 115 (left) and 230 (right) kbaud. The blue dots show the PER measured in the corridor, where artificial illumination is always present, whilst red dots are instead reporting calibration of PER, performed in a laboratory with artificial lights turned off. Every point shown in Fig. \ref{fig:PER_vs_amp} is the average of three consecutive measurements. Vertical error bars (masked by symbols size in figure) are given by the maximum deviation. Horizontal error bars correspond to an error of $\pm$1\,mV ($\pm$0.5\,mV) on amplitude measurements in the lights-on (lights-off) configuration. Solid curves represent a theoretical fit to data performed through Eq.\,\eqref{eq.Qfunc}. In the fit, we consider the experimental errors as weights, and we leave $T$ and $\sigma$ as free parameters. The shaded areas represent the error associated to the fit procedure, where experimental errors are used as weights.
As Fig.\,\ref{fig:PER_vs_amp} clearly highlights, few mVs are sufficient for our RX stage to grant an error-free transmission, and the lower 115 kbaud has higher performances at a given amplitude, as expected. Given the extremely small values of received intensity, our analysis also shows that the VLC link quality is negatively affected by the residual effects of 100\,Hz illumination after the AC decoupling stage, which requires stronger VLC signal at RX stage to grant the same PER with respect to the dark case. This can be easily understood, as for low amplitudes the residual 100\,Hz slow fluctuations in the VLC signal can periodically shift the signal with respect to the pre-set comparator threshold detection, hence inducing periodic reading errors in the RX board. This observation confirms our expectations of outdoor scenarios to be less critical (in absence of bad weather conditions) as compared to the indoor case.
Importantly enough, hence, we will use the lights-off calibration in the rest of the paper, as it provides for a more accurate estimation of PER performances of our system when used in outdoor scenarios, where no artificial 100\,Hz background effects are expected. We also note that by using a more powerful digital signal processor such as FPGA would easily solve such slow signal fluctuations via quasi-real-time signal analysis, at the probable expense of the ultra-short latency capabilities of our system \cite{our_paper2}.

\begin{figure}[]
\centering\includegraphics[width=\columnwidth]{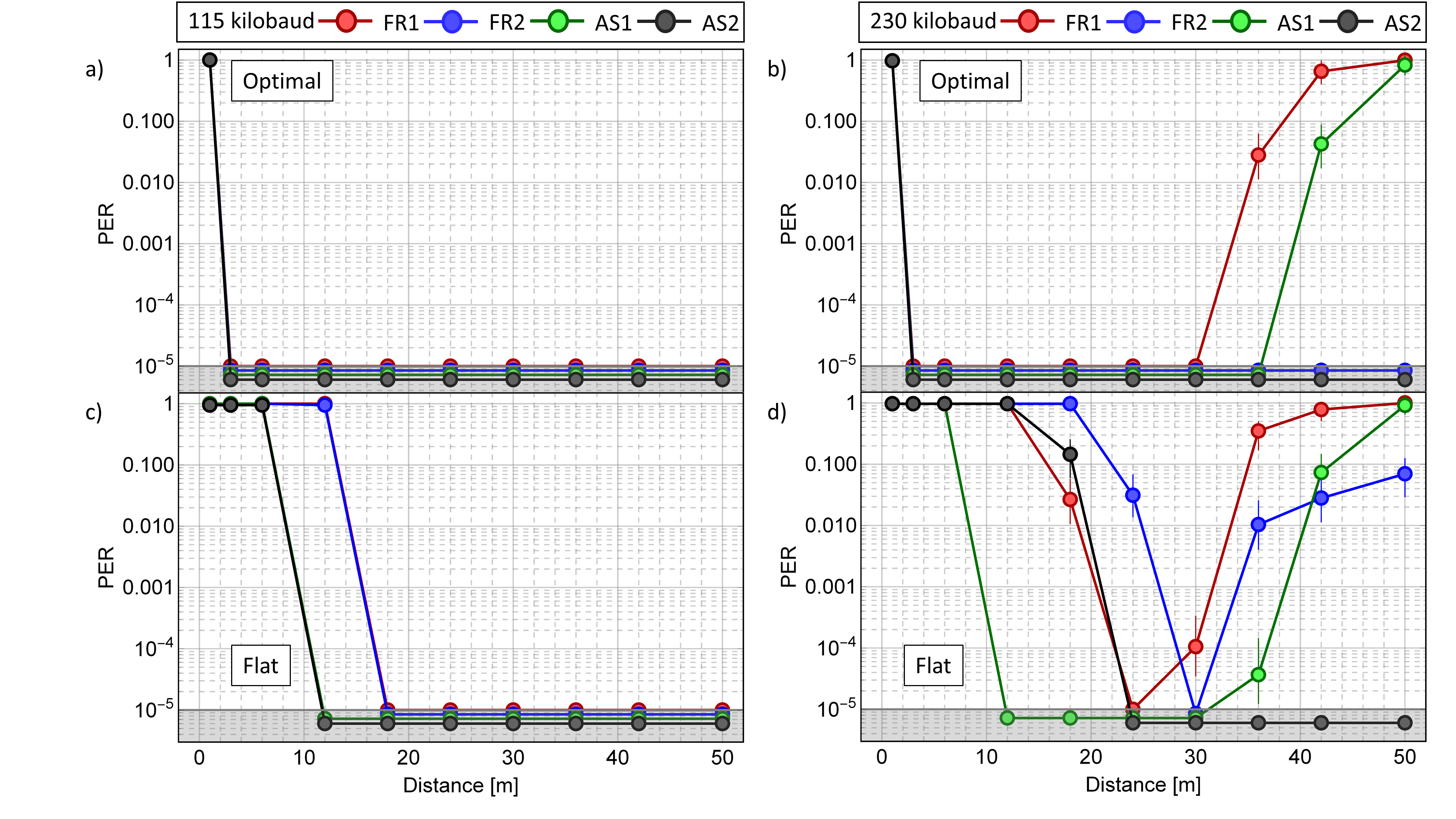}
\caption{PER measured as a function of distance for 1.5 m side configuration for all lenses. Two baud rates of 115 kbaud (left) and 230 kbaud (right) are used for data transmission. The upper panels show the optimal configuration while the bottom panels show the flat configuration.}
\label{fig:PER_vs_distance_outdoor}
\end{figure}


By combining results shown in Figs.\,\ref{fig:Amp_vs_distance}-\ref{fig:PER_vs_amp}, we can now assess the performances of our system in terms of PER for various condensing elements and positions across the measurement grid. 
Fig.\,\ref{fig:PER_vs_distance_outdoor} reports the PER measured as a function of distance for the optimal and flat RX orientations in the side configuration where the photodetector is placed 1.5\,m aside the traffic light axis (corresponding to the realistic scenario of a car approaching the crossing in the rigthmost road lane with no auto-tracking mechanism for RX orientation). Results are shown for all of the available lenses. For 115\,kbaud (left panels), in optimal configuration (panel a)) an error-free communications (PER$<10^{-5}$) is remarkably achieved up to 50\,m and down to 3\,m with \textit{all} lenses. As expected from results of Fig.\,\ref{fig:Amp_vs_distance}-d, however, AS1 and AS2 lenses are the most suited at short distances in the side-flat case (panel c)), granting efficient VLC links for distances not shorter than 12 m. The situation gets more complex at higher baudrates (230 kbaud), as larger signal amplitudes are required to grant the same SNR. However, in the optimal case (panel b)), 2" lenses still grant a nearly error-free link from the whole range 3--50\,m, the 1" lenses suffering from worse performances at distances $>30$\,m due to lower optical gain. In the most demanding side-flat configuration (panel d)) the AS1 again represents the only  suited condenser in the medium-short distances arange 12--30 m, while AS2 is the only lens granting error-free connections at larger distances up to 50\,m. 

In this extreme case of side-flat, higher baudrate configuration it is evident from our results that the ultimate optical performances of the condensing element are a key feature, and aspheric elements are outperforming the corresponding Fresnel ones in both short and long range connections.
Despite no lens allows for error-free communication at all distances in this realistic case, we highlight that by combining a 115k baudrate to an AS1 or AS2 condenser our system can establish an error-free connection in the range 12--50\,m and a PER$<10^{-3}$ for all distance above 10\,m. We remark that, following our recent analysis reported in \cite{our_paper2}, we expect sub-ms statistical latency values for short message broadcasting in the whole PER$<10^{-3}$ region. 
Our measurements also highlight how, in real I2V ITS VLC applications in proximity of road crossings, the most critical region appears to be the short-distance one ($<10$\,m), where the connection between a traffic light and a vehicle appears to be particularly hard for any of the examined lenses due to the high FOV required.

\subsubsection{Measuring effects of misalignment on telecom performances}\label{subsub:FOV} 
\begin{figure}[]
\centering
\includegraphics[width=\columnwidth]{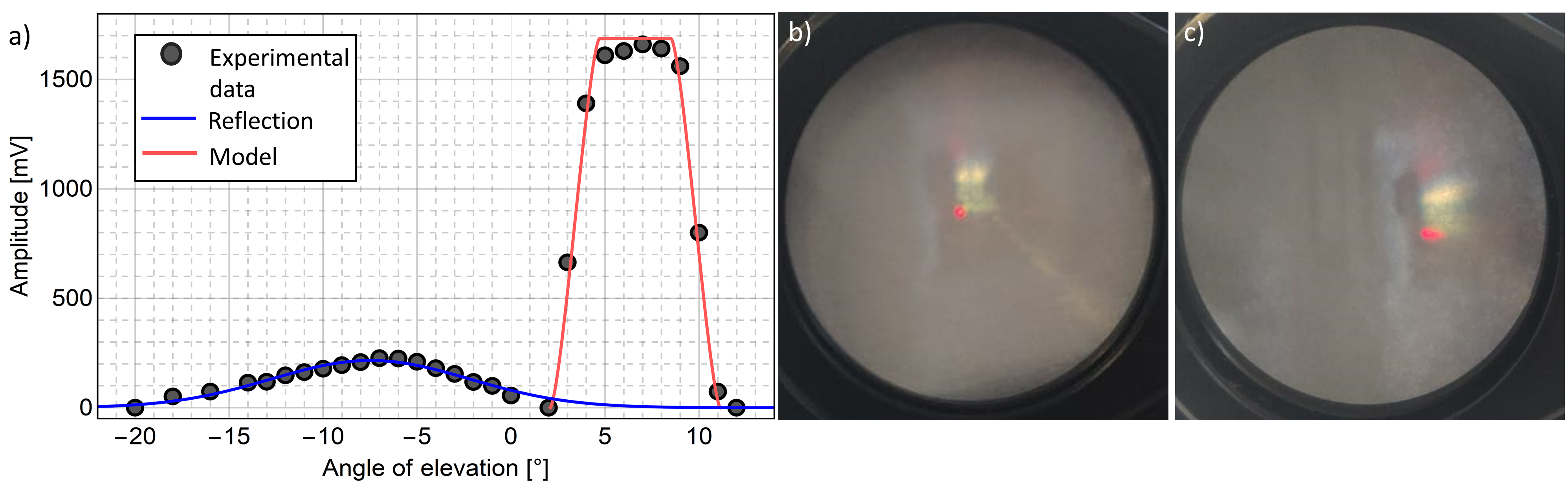}
\caption{a) Typical amplitude measured at RX stage as a function of the angle of elevation $\alpha$. The black dots show the data, with error bars on both angle and amplitude smaller than symbols size. The blue line evaluates the floor reflection contribution with a Gaussian fit, whilst the red line is our fit to direct amplitude values using Equation \ref{eq:model_eq}. Panels b) and c) show an image of traffic light lamp taken in the focal plane of the condenser lens (FR2 in this case) for both on-axis (a) and off-axis (b) image configurations. The off-axis configuration suffers from an evident coma effect, with a pronounced tail on the right side of image. Both panels refer to data taken at 18\,m in the front case.}
\label{fig:Reflection}
\end{figure}
\begin{figure}[]
\centering\includegraphics[width=\columnwidth]{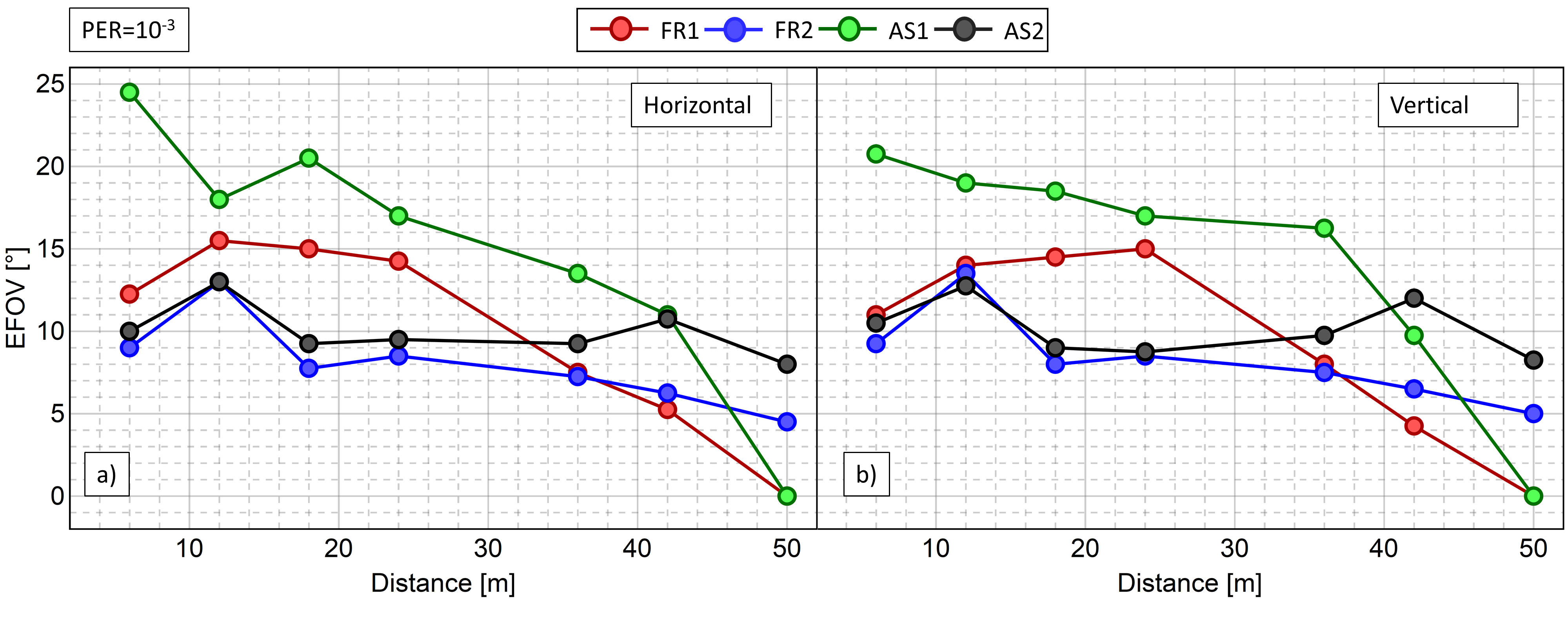}
\caption{Experimental characterization of EFOV vs. distance for 115 kbaud data transmission for all lenses. Error bars on both angles and distances are smaller than symbol size. $PER=1\times10^{-3}$ is taken as a threshold value (see Sec.\ref{sub:trans_performances}).}
\label{fig:EFOV}
\end{figure}

Effects of relative misalignment between TX and RX in performances of our VLC prototype can be quantified through an experimental assessment of the EFOV. To this scope, we placed our RX stage on precision rotation platforms, allowing us to record the received signal amplitude as a function of both $\phi_H$ and $\phi_V$ angles, with 0.5\degree\, resolution, for various distances and lens sets.  In performing such an analysis, which requires large scans of the angular position of RX, we could observe several non-trivial contributions, arising from  reflections of the traffic light signal on either walls, floor or ceiling. The full characterization of such reflections on the VLC channel in a real urban scenario is definitely an important aspect and is subject of future works, and we highlight here a method for efficient isolation (and eventual quantification) of reflections contributions in post-analysis.

As an example, Fig.\,\ref{fig:Reflection}-a reports the measured amplitude as $\phi_V$ is scanned on a certain point on the grid in the optimal configuration, for AS2 lens. Data highlight a strong contribution of reflections (left) at low and negative elevation angles, which partially overlaps with the mean peak (right) coming from the direct VLC signal. In order to isolate the contribution reflections from the direct signal feature (which is used to retrieve the EFOV), we fit our data to a combined function, which is the sum of the amplitude function $S(\alpha, \beta, d)$ given by Eq.\,\ref{eq:model_eq} and of a gaussian function where all of the four parameters (width, height, baseline, center) are free. In order to heuristically account for aberrations effect, the main being coma (as Fig.\,\ref{fig:Reflection}-b shows), and defocusing (the RX doesn't have an auto-focus system and the lens is kept at a constant distance equal to $f$ from photodiode), we embed such effects in the radius of  image appearing in Eq.\,\ref{eq:model_eq}, by considering it as a free parameter. This combined fit procedure provides an excellent agreement with data, and allows us to quantitatively isolate the contribution of reflections (blue curve) and to retrieve a reliable estimation for $S(\alpha, \beta, d)$ (red line) from each dataset on the measurement grid even for small amplitude values.

 \begin{figure}[]
\centering
\includegraphics[width=\columnwidth]{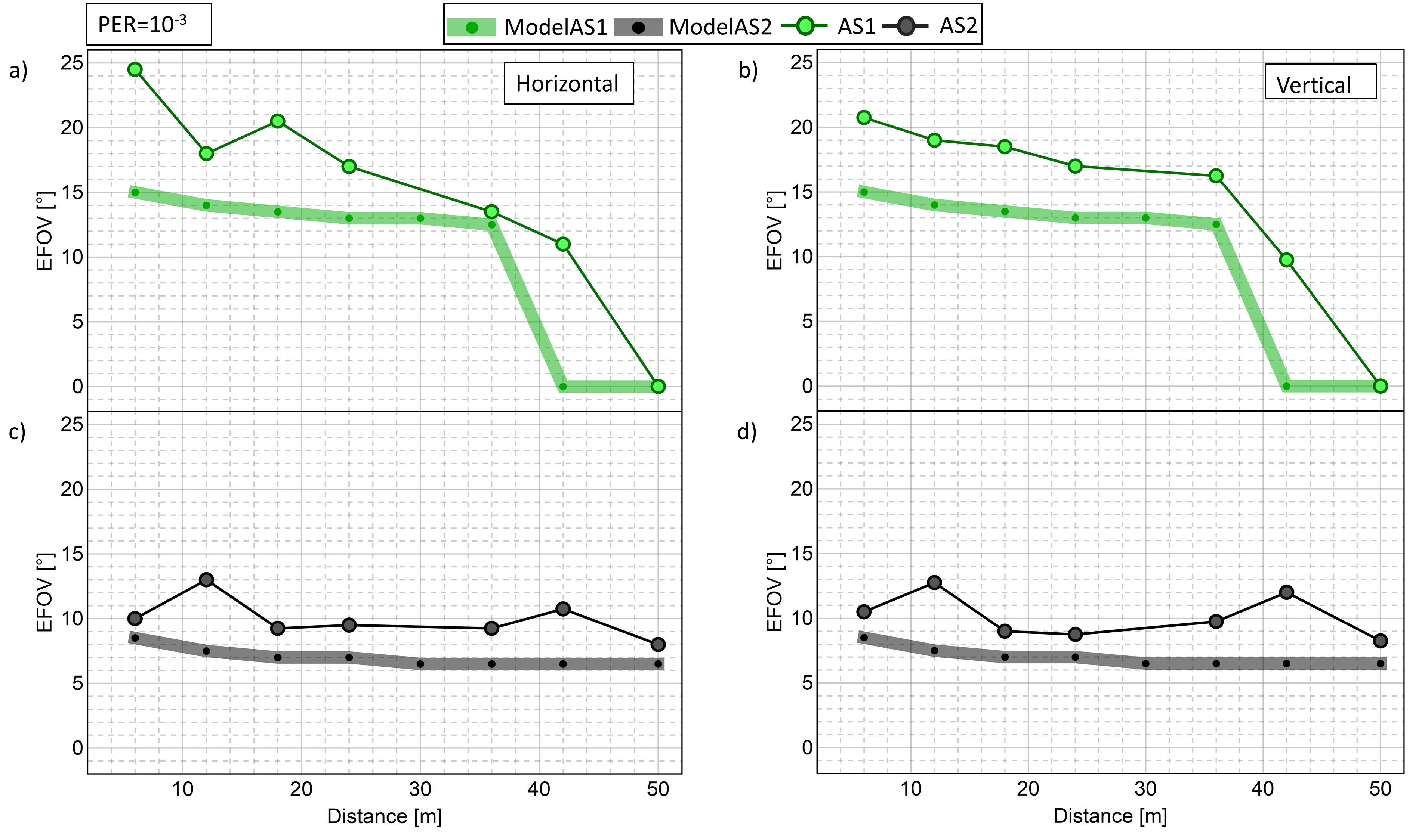}
\caption{Comparison between measured EFOV (symbols) and model predictions (shaded areas) for AS1 ans AS2 lenses as a function of distance for a baudrate of 115 kbaud. The vertical error bars are smaller than symbol size. The upper and lower bounds of model, represented by the extension of shaded areas, are calculated by considering $\pm$ 0.5 mV as uncertainty on measured amplitude in model predictions.}
\label{fig:EFOV_exp_theo}
\end{figure}

Once the amplitude $S$ id extracted from data, we can retrieve the PER as a function of  $\phi_{H,V}$ exploiting the calibration procedure reported in previous sections, and then provide for an experimental characterization of EFOV (see Sec.\,\ref{sec:model}) as a function of distance for different lenses. Fig.\,\ref{fig:EFOV} reports such analysis for both vertical and horizontal EFOV in the range 6--50 m for all lenses. In case of AS1 lens, and for a baudrate of 115k, the system remarkably features EFOVs as high as 25\degree, achieved for short distances where the intensity is higher. This large acceptance angle is what grants the possibility to establish a VLC connection for distances as low as 3\,m (see Fig.\,\ref{fig:EFOV}), where RX-TX angles are very large. The short-focal AS1 lens outperforms all the other lenses except for very large distances, where the small diameter reduces the collected optical power, so that the PER=$10^{-3}$ threshold is harder to be achieved when compared to larger lenses. Interestingly, the 2" lenses, and particularly the AS2 lens, show instead a rather constant behaviour with EFOV attesting around 10\degree, in the whole distances range. This is given by a combined effect of reduced EFOV at short distances, connected to the limited AFOV due to the longer focal length, which is in turn compensated by the large diameter of 2" lenses, limiting the long-distance intensity drop. 

Fig.\,\ref{fig:EFOV_exp_theo} reports a comparison between experimental data on EFOV for AS1 and AS2 (full symbols) and predictions of our model given in Sec.\,\ref{sec:model} (shaded areas). The error on model (width of shaded areas) is estimated by assuming $\pm 0.5$ mV as experimental uncertainty on amplitude values and calculating the two corresponding model values for EFOV. We see that the global trend is fully caught by model at both short and large distances, and the quantitative agreement is  good, with a global underestimation of model with respect to data, leading to a modest discrepancy on the whole distance range (except than in the narrow transition region towards EFOV = 0 in the caso of AS1). This discrepancy can be attributed to several factors. First, as the very steep trend in PER vs amplitude given by the calibration curves of Fig.\, \ref{fig:PER_vs_distance_outdoor} confirms, the performances of our setup in terms of PER heavily depend on amplitude around the threshold, and differences of few mVs can alter the PER by orders of magnitudes. Second, as Fig.\,\ref{fig:Model_geometry} shows, the effect of angular rotation on received amplitude is very drastic, leading to a steep decrease of the amplitude in the intermediate region of model (\ref{eq:model_eq}). Hence, the combination of these behaviours tells us that the data-model discrepancy is actually is actually very low in absolute terms of amplitude, as it lays in the few-mV range. On the other side, the occurrence of slight deviations from the smooth trend in the measured data (e.g. for $d=12$ m in both panels a) and b)) is likely due to residual imperfections in the reflections isolation procedure for particular positions on the grid, where the reflection component is more overlapped to the direct signal and hence harder to be deconvolved.  
A full characterization of reflections effects in outdoor realistic scenarios is an important research topic which will be covered in future experiments.

\section{Conclusions}\label{sec:conclusions}

In this paper a thorough characterization of a VLC system for ITS I2V applications is reported. The paper analyzes both the optical and the telecom performances with a broad set of optical condensers in a realistic set up. A regular traffic light, enabled for VLC transmission, is used as source and a receiver composed by an optical condenser and a dedicated amplified photodiode stage.  We performed our experimental measurements for distances in the range 3--50\,m in terms of both PER and EFOV. The results show several nontrivial behaviors for different lens sets as a function of position on the measurement grid. The results also show that some configurations are suitable for an efficient transmission with PER better than $10^{-5}$ up to 50\,m, with EFOVs higher than 10\degree\,. We highlight how in such applications the most critical region is in the 3-10\,m range, due to the limited FOV of the optics and the large relative angles involved between TX and RX stages. In addition, in our paper we also provide a theoretical model which is an angle-dependent extension of the one reported in \cite{2019arXiv190505019C}, for both the signal intensity and EFOV as a function of several parameters, such as distance, RX orientation and focal length. The proposed model could be useful in predicting the performance of a VLC system in real scenarios, where angle-dependent misalignment effects are unavoidable. 

To our best knowledge, this is the first time that an EFOV analysis for VLC systems for ITS is reported and detailed. Actually, it has more general implications with respect to the pristine I2V case mentioned here, as it quantifies an intrinsic performances of a VLC RX stage, composed by an optical condenser of a given focal length and a photodiode of a give size. Our results could be very relevant in the future to assess a most suited solution in terms of acceptance angle when designing a VLC system. Depending on the specific VLC application, indeed, one can choose to privilege short focals with large acceptance angles (when a strong directivity of the channel is not required, e.g., in typical indoor data dissemination), or target a long cast in transmission range at the expense of a reduced acceptance angle, in applications where the VLC channel directivity is rather a key factor (as in I2V traffic light-to-car and V2V car-to-car applications).
\color{black}

\section*{Acknowledgements}
The authors warmly thank F.\,S.\,Cataliotti for a careful reading of the paper and for precious discussions. This work has been carried out under the financial support of PON MIUR 2017 "OK-INSAID" and Progetto Premiale MIUR FOE 2015 "OpenLab".


\bibliographystyle{IEEEtran}
\bibliography{IEEEabrv,Bibliography}

\end{document}